\documentclass[structabstract]{aa}
\usepackage{graphicx}
\usepackage{txfonts}
\usepackage{latexsym}
\usepackage{bm}

\begin{document}

\title{No quantum gravity signature from the farthest quasars}

\author{F.~Tamburini~\inst{1}
\and
C.~Cuofano~\inst{2} 
\and
M.~Della~Valle~\inst{3,4}
\and 
R.~Gilmozzi~\inst{5} }

\institute{
Dipartimento di Astronomia, Universit\`{a} di Padova, 
vicolo dell'Osservatorio 3, 35122 Padova, Italy \\
\email{fabrizio.tamburini@unipd.it}
\and
Dipartimento di Fisica, University of Ferrara.\\
\and
INAF - Osservatorio Astronomico di Capodimonte, salita Moiariello 16, 80131 Napoli, Italy.\\
\and
International Center for Relativistic Astrophysics Network, Pescara, Italy.\\
\and
European Southern Observatory, Garching bei M\"unchen, Germany
}

\date{Received ...; accepted ...}

\abstract
 % context heading (optional)
{Strings and other alternative theories describing the quantum properties of space-time suggest that space-time could present a foamy structure and also that, in certain cases, quantum gravity (QG) may manifest at energies much below the Planck scale. One of the observable effects could be the degradation of the diffraction images of distant sources.}
% aims heading (mandatory)
{We searched for this degradation effect, caused by QG fluctuations, in the light of the farthest quasars (QSOs) observed by the Hubble Space Telescope with the aim of setting new limits on the fluctuations of the space-time foam and QG models.}
% methods heading (mandatory)
{We developed a software that estimates and compares the phase variation in the interference patterns of the high-redshift QSOs, taken from the snapshot survey of HST-SDSS, with those of stars that are expected to not be affected by QG effects. We used a two-parameter function to determine, for each test star and QSO, the maximum of the diffraction pattern and to calculate the Strehl ratio.}
% results heading (mandatory)
{Our results go far beyond those already present in the literature. By adopting the most conservative approach where the correction terms, that describe the possibility for space-time fluctuations cumulating across long distances and partially compensate for the effects of the phase variations, are taken into account. We exclude the random walk model and most of the holographic models of the space-time foam. Without considering these correction terms, all the main QG scenarios are excluded. Finally, our results show the absence of any directional dependence of QG effects and the validity of the cosmological principle with an independent method; that is, viewed on a large scale, the properties of the Universe are the same for all observers, including the effects of space-time fluctuations.}
% conclusions heading (optional)
{}


\titlerunning{No QG signatures from QSOs}

\maketitle

\section{Introduction}

Quantum gravity (QG) can be considered the Holy Grail of modern physics. Because of the general Heisenberg's uncertainty principle, in most of the approaches to the unification of gravitation and quantum physics, the classical geometrical concept of a continuous and smooth space-time support of general relativity is replaced, at very short distances, by a foamy geometric scenario where fluctuations on the order of the Planck length $l_p\simeq~1.616\times10^{-35}~\mathrm{m}$  occur at certain characteristic frequencies that are related to the Planck time $t_p\simeq~5.391\times10^{-44}~\mathrm{s}$. For the sake of simplicity, here we will not consider in this discussion more exotic formulations of Heisenberg's principle, whici can be found in the vast literature of quantum gravity.

Without going into too much detail, each model of QG can be naively characterized by its peculiar space-time and energy fluctuation scale, which may occur also at frequencies that are different or even much lower than those expected at Planck scales, $1/t_p$. The reader can find quite a vast literature describing all the possible scenarios and behavior of gravity on quantum scales, because up to now there has been no theory of QG.
This scale is thought to be accompanied by the occurrence of new physical phenomena and macroscopically observable effects, such as the production of mini-black holes at energies much lower than the classical Planck scale, like those expected in the large hadron collider 
(LHC, 2010; Dvali \& Sibiryakov, 2008; Koch et al., 2005; Hossenfelder, 2006; Hossenfelder, 2002).
%\cite{lhc,dva08,koc05,hos06,hos02}. 
Other models even describe the possibility of loosing of phase coherence in the light of distant sources, 
such as the farthest supernovae (SNe) and distant quasars (QSOs). Light, when propagating in the vacuum, will be affected by the fuzzy structure of space-time, with the result of progressively 
blurring the diffraction images of those sources (Amelino-Camelia, 1994; Amelino-Camelia, 2000; Ng \& van Dam, 2000; Ng \& van Dam, 1994; Ng, 2003; Christiansen et al., 2006).
%\cite{ame94,ame00,ngv00,ngv94,ng03,chr06}.

Naively, following the already cited literature, one expects that a photon with energy $E=h\nu~$ is 
affected by an energy indetermination on the order of $\delta E/E \sim \delta p/p \sim (E/E_p)^\alpha$, 
where  $\alpha$ is the \textit{characteristic exponent} of the power--law distribution of a 
phenomenological model that describes the energy fluctuations induced in any quanta by QG effects 
because of the ``grainy'' structure of space-time. The quantity $E_p =h/t_P$ is Planck's energy. 
The energy-momentum relationship at the lowest perturbation order is a function of the two parameters 
$a_0$ and $\alpha$,
\begin{equation}
p^2= E^2[1\pm a_0(E/E_P)^\alpha]. 
\end{equation}

If this ``grainy'' structure of space-time emerges on time scales of $t^\ast\geq t_P$, any time interval 
can be determined with an uncertainty $t\geq t^\ast$ and intrinsic standard deviation derived from the 
following relationship, $\sigma_t/t  = f(t_p/t)$, where $f \geq 1$ for $t \leq t_P$. When, instead, 
$t \gg t_P$, one obtains $f=(t_p/t)(a_0+a_1(t_p/t)+a_2(t_p/t)^2+ ...) \approx a_0(t_p/t) \ll 1$, 
where $a_0$ is the lowest order amplitude of the quantum fluctuations in Planck units.
This means that this foamy structure of space-time induces a random dispersion of the photon 
arrival time of a bunch of photons emitted by a source. 
It is possible to reveal this effect by analyzing the phase coherence of the light of distant sources,
as described by \cite{lh03} and \cite{rtg03}.
Equivalently, this dispersion is said to generate two differently fluctuating wave velocities that may 
systematically cumulate during the journey of each photon. The wavelength $\lambda$ of the traveling 
photon naturally defines the minimum and characteristic scale length on which physical quantities such 
as, the phase velocity ($v_p$) and the group velocity ($v_g$), and their respective dispersion relationships 
can be defined. 

The phase of the electromagnetic wave, associated with the photon propagation, after having crossed 
the distance $\lambda$ in the time $t= \lambda /v_g$, becomes $\phi=2\pi v_p/v_g$. 
\cite{lh03} and \cite{nlo01} claim that QG fluctuations may induce well-defined phase indeterminations
such that an electromagnetic wave, traveling a long distance $L$, may cumulate a phase variation on the order of 
\begin{eqnarray}
(\Delta \phi_{QG})_{min} &\sim& 2 \pi a_0 \frac{l_p^\alpha L^{1-\alpha}}{\lambda} \nonumber \\
&\sim& 2 \pi \frac {L}{\lambda} \left[ 1\pm \sqrt{2} \alpha \left( \frac{hc}{\lambda E_P} \right)^\alpha \right],
\end{eqnarray} 
if the emission and detection wavelengths remain equal during the photon propagation. The parameter $c$ 
is the speed of light and $a_0\sim 1$ (Ng, 2003).
If, instead, one also considers  the effect of cosmological redshift $z$, a larger phase dispersion is obtained, as proposed by  \cite{str}
\begin{eqnarray}
&&(\Delta \phi_{QG})_z \sim 2 \pi a_0 \frac{l_p^\alpha}{\lambda}\frac{(1-\alpha)c}{H_0q_0} \nonumber \\
&&\times \int_0^z(1+z)L^{-\alpha}
\left[1-\frac{1-q_0}{\sqrt{1+2q_0z}}\right]dz ,
\label{eq3}
\end{eqnarray}
where $H_0$ is the Hubble constant, $q_0$ the deceleration parameter of  the Universe, 
and the luminosity distance is given by
\begin{eqnarray}
L=\frac{c}{H_0 q_0^2}\left[q_0 z -(1-q_0)(\sqrt{1+2q_0 z}-1)\right] \,\,.
\end{eqnarray}

This oversimplified phenomenological description of QG models the foamy structure of space-time with 
only the two parameters $a_0$ and $\alpha$ at the first order in amplitude fluctuations.
By varying $\alpha$, one obtains the three main scenarios of QG.
The random walk model has $\alpha=1/2$ and random perturbations of space-time add incoherently with 
a square root dependence in the simplest case, or even following a more general distribution 
function. Examples are the works by \cite{ame00} and \cite{ngv94}.
The holographic model, which is consistent with Beckenstein-Hawking's holographic principle and black hole 
entropy, has $\alpha=2/3$ ('t Hooft, 2003). %\cite{tho93}.  
The value $\alpha=1$ describes the standard model of QG in which the gravitational field fluctuations 
are close to the Planck time (Smolin, 2003; Rovelli, 2004).
% \cite{smo03,rov04}.

Those effects can in principle be detected by analyzing possible modifications of the interference 
fringe structures using stellar interferometry. Equivalently, the cumulated phase 
variation can also be detected  by imaging and analyzing, with a diffraction-limited 
telescope, the ring structure of Airy disks of cosmological point-like sources (Ragazzoni et al., 2003).%\cite{rtg03}. 
Airy rings are still observed if the cumulated phase is limited to the value $\delta \phi \leq 2 \pi$. 

Previous results, based on data from stars and extragalactic sources (Ng \& van Dam, 1994; Lieu \& Hillman, 2003), %\cite{ngv94,lh03}, 
seemed to rule out both the random walk model and the holographic principle and to put severe limits on the standard model. Similarly, the standard model also appeared to be ruled out by data on PKS1413+35 (z=0.247) and by the detection of the Airy rings from SN 1994D (z= 0.001494) (Ragazzoni et al., 2003). %\cite{rtg03}. 
However, in these conclusions the authors did not consider the more 
recent (Ng, 2003)  possibility that the space-time fluctuations may cumulate across long distances and partially compensate for the effects of the phase variations. In this case the correction factor is $(L/ \lambda)^{-\alpha}$ and
only the random walk model could be excluded with the data available so far.
A more sensitive analysis performed by \cite{str}, about the Strehl ratios of high-redshift 
QSOs, pointed out a possible deterioration of the interference structures due to the space time foam.

In this paper we resume the test on QSOs with a new approach to the data analysis and we put new 
upper limits on the QG fluctuations by analyzing the images of distant QSOs, observed by the Hubble Space Telescope (HST) with a software ad-hoc developed for this purpose.

With our estimates, if we neglect those possible cumulating phase effects, all the models of space-time foam would be immediately ruled out, thereby posing strict limits on the standard model. 
However, in our paper we use a more conservative approach, in which the space-time fluctuations are expected to compensate for their variations  owing to their ``averaging'' and ``smoothing'' during the propagation of the light beam, causing the lack of detection of space-time fluctuations.

\section{Data analysis}

We searched for QG effects on the interference patterns of the high-redshift QSOs taken from the archival HST ACS snapshot survey of Sloan  Digital Sky Survey (SDSS) sources (Proposal: 9472, PI: Strauss) and described by \cite{ric06}. The dataset consists of 157 sources located in the redshift interval $4 < z < 5.4$ and acquired with filter  F775W ($\lambda_\mathrm{c}=776.4$ nm, bandwidth $\Delta \lambda=152.8$ nm). Four other sources, located at redshifts $5.7 < z < 6.3$, were instead taken with the filter F850LP ($\lambda_\mathrm{c}=944.5$ nm, bandwidth $\Delta \lambda=122.9$ nm). The detector used in the observations was the High Resolution Camera (HRC), which has a pixel scale of $0.0246$ arcseconds.
We used in our analysis the \textit{drizzled} (DRZ) images that were corrected for geometric effects. The DRZ files were all reduced by adopting the standard procedure, including the overscan-, bias-, and dark-corrections and the flat-field and the photometrical calibration.

The luminosity distances of the QSOs, $L$, were calculated by assuming a concordance Universe model described in \cite{kra02}: $H_0=72$ km/s/Mpc, 
($\Omega_M, \Omega_\Lambda$)=(0.3, 0.7), and also by taking the small effects on $\lambda$ 
due to the expansion of the Universe into account. 

To determine the QG effects through the progressive image degradation of a distant 
source, we estimated the Strehl ratio $S$ of the source  for a given wavelength $\lambda$
through the empirical phase variation
\begin{equation} 
\Delta \phi \approx \frac{\lambda}{D} \sqrt{-ln (S)}
\end{equation}
where $D=2.4$ m is the telescope diameter.
The Strehl ratio $S$ is defined as the ratio between the actual observed peak intensity 
($\mathrm{Max})_\mathrm{source}$ and what is theoretically expected from an unaberrated telescope 
($\mathrm{Max})_\mathrm{th}$
\begin{equation} 
S=\frac{\mathrm{(Max)}_\mathrm{source}}{\mathrm{(Max)}_\mathrm{th}}\sim \exp 
\left[ -\left(\Delta \phi \frac {D}\lambda \right)^2\right].
\end{equation}
From the Strehl ratio $S$ we obtain the total phase variation $\Delta \phi$ that for QSOs can be written as
\begin{equation}
 \Delta \phi(z)_\mathrm{QSO} = \Delta\phi_\mathrm{ab}+\Delta\phi(z)_\mathrm{size}+\Delta\phi(z)_\mathrm{QG}+\Delta\phi(z)_\mathrm{lens} \, ,
\label{eq5}
\end{equation}
where $\Delta \phi_\mathrm{ab}$ is the phase variation due to the aberrations 
present in the optical path of HST, $\Delta \phi_\mathrm{QG}$ those due to QG,
$\Delta \phi_\mathrm{size}$ to the resolvability of QSOs and/or their host galaxies, 
and $\Delta\phi(z)_\mathrm{lens}$ to the gravitational lensing. The  phase variation due to gravitational lensing, averaged in all the exposures of the dataset, is expected to behave as an increase in the blurring already caused by the QG effects, so it cannot, in principle, be easily separated from the latter ones except in special cases. The presence of spikes, caustics, and other typical effects of microlensing were not been observed in our sample. Because of this, the detection of QG effects would be overestimated, because unavoidably entangled with gravitational lensing and optical aberration effects.

To reduce the blurring caused by the aberrations of the telescope, we modeled this image degradation effect 
by analyzing a sample of pictures of stars.
The effects of QG are negligible at non cosmological distances and $\Delta\phi_\mathrm{star}\simeq\Delta\phi_\mathrm{ab}$.

In the case of the QSOs, we now consider the quantity
\begin{eqnarray}
(\Delta\phi(z)_{QSO})_\mathrm{Corrected} &=&  
\Delta \phi(z)_\mathrm{QSO}-\langle\Delta \phi\rangle_\mathrm{star} \nonumber
\\
&\simeq& \Delta\phi(z)_\mathrm{size}+\Delta\phi(z)_\mathrm{QG}+\Delta\phi(z)_\mathrm{lens} \pm \delta\phi_{star}.
\label{eq6}
\end{eqnarray}
To adopt a more conservative approach, we use the averaged value of the phase variation $\langle\Delta \phi\rangle_\mathrm{star}$ instead of the dispersion of $\Delta\phi_\mathrm{star}$ values found in the sample of the stars analyzed by us to calibrate our procedure.
The quantity $\delta\phi_{stars}$ includes the errors due to the data analysis, the different positions on the detector and the different colors of the stars.  The indeterminant $\delta\phi_{color}\lesssim 10^{-8}$~rad represents the error due to the different colors found between the stars and QSOs. This quantity has been estimated by analyzing the PSFs generated by the Tiny Tim software.

Our approach is different from the procedure applied by \cite{str}, in which the Strehl ratios of QSOs were not corrected but only compared to an appropriate PSF generated for each of the QSOs by the Tiny Tim software. In fact, we did not assume any particular PSF model in our investigation, and we corrected the phase variation of QSOs by using the correction terms derived empirically from galactic stars in Eq.(\ref{eq6}).

For our analysis we developed a dedicated software to determine the position and the maximum of QSOs and stars.
For each source imaged by the ACS camera, we assumed that the central region of the interference structure has an elliptical symmetry. The position and the value of the maximum of the light pattern are determined in a region of $3\times 3$ pixels. The light intensity distribution is then interpolated by a 2-D function

\begin{eqnarray}
&F&(x,y,[mx,cx,cy,\epsilon,\theta, \sigma_1, \sigma_2])  =  \frac{mx}{\sigma_1+\sigma_2}\nonumber \\ 
&\times& \sum^2_{n=1}\sigma_{3-n}\exp\left[-\frac{x_1^2(x,\cdots)-\epsilon x_1^2(x,\cdots) 
+ y_1^2(x,\cdots)}{2\sigma_n^2(1-\epsilon)}\right]
\label{methods}
\end{eqnarray}
where the pair $(cx,cy)$ identifies the position of the maximum $(mx)$ in the Cartesian coordinate system, the origin of which is placed on the center of the brighter pixel. The quantity $\epsilon$ is the eccentricity of the ellipse, and
$\theta$ is the rotation angle between the axes of the ellipse and the cartesian axes. The parameters 
$\sigma_1$ and $\sigma_2$ modify the shape of the function $F$, and we 
write the argument of Eq.~\ref{methods} as
\begin{eqnarray}
x_1(x,y,cx,cy,\theta)&=&(x-cx)\,\mathtt{cos}\,\theta-(y-cy)\,\mathtt{sin}\,\theta \nonumber \\
y_1(x,y,cx,cy,\theta)&=&(x-cx)\,\mathtt{sin}\,\theta+(y-cy)\,\mathtt{cos}\,\theta \,\,.
\end{eqnarray}

\begin{figure}
\centering
\includegraphics[width=9 cm, keepaspectratio]{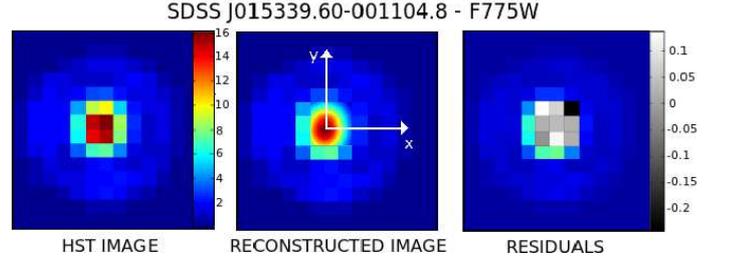}
\caption{Left: HST image of the QSO QSO 0153-0011, IAU name SDSS J015339.60-001104.8 located at $z=4.205$.
The field of view is $15\times 15$ pixels. 
Center: we show the central region ($3\times 3$ pixels) reconstructed using the function $F$ (Eq.\ref{methods}). 
We plot the cartesian axes centered on the maximum of the intensity pattern. Here $\epsilon=0.23$ and $\alpha=0.14$~rad. 
Each pixel of the analyzed area is divided into $15\times 15$ sub--pixels. 
Right: we show the residuals of the central analyzed area ($3\times 3$ pixels).}
\label{fig1}
\end{figure}

\begin{figure}
\centering
\includegraphics[width=9 cm, keepaspectratio]{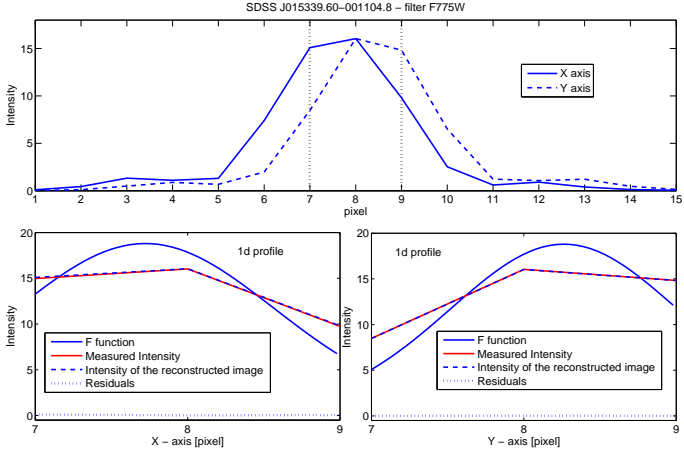}
\caption{Top: 1d cut along the $x$--axis (solid line) and the $y$--axis (dashed line) 
of the measured intensity of QSO SDSS J015339.60-001104.8. 
We analyzed the area within the two dotted lines. Bottom: we show the measured intensity (red line),
the interpolated function $F$ (solid blue line), the intensity of the reconstructed image (dashed blu line), and the residuals (dotted blue line)
of the 1-d cut along the $x$--axis (left) and the $y$--axis (right) of the central area ($3\times 3$ pixels).}
\label{fig2}
\end{figure}

With a recursive procedure, we tune the values of the parameters $mx$, $cx$, $cy$, $\epsilon$, $\theta$,  $\sigma_1$,  $\sigma_2$ of the 2D function, $F$, until minimizing the difference between the measured and the fitted intensity of each pixel, on a $3\times 3$ matrix situated around the brightest pixel. In the calculations, each pixel is divided into $15 \times 15$ subpixels, and for each of these divisions the procedure determines the intensity as a function of the position in the detector. 
At the end, we obtained both the position of the center and the intensity of the light pattern in the $3 \times 3$ area with a precision of $15 \times 15$ division for each pixel (see Figs.\ref{fig1} and \ref{fig2}). 
We found deviations between the fitted and measured intensities to be less than $1\%$. 
It is important to point out that we considered only the light distribution inside the region delimited by the inner Airy diffraction ring. 
The values of the maxima of stars and QSOs are normalized with respect to the flux inside an area, centered on the brightest pixel, with a radius of $2$ arcseconds. An example is reported in Figs. 1 and 2 with the QSO QSO SDSS J015339.60-001104.8 (IAU) (z=4.205).

%% Fig1

%% Fig2

%Fig3
\begin{figure}
\centering
\includegraphics[width=8.1 cm, keepaspectratio]{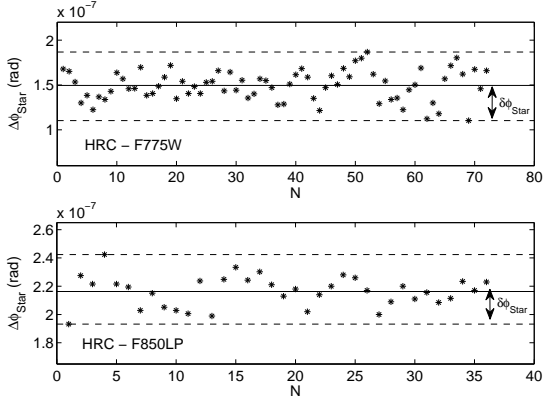}
\caption{Phase variation of 72 stars imaged with the High Resolution Camera with the filter F775W
(upper panel) and 36 stars with the filter F850LP (bottom panel). 
We plot also the averaged value $\langle\Delta\phi_{star}\rangle$ (solid line) and the error
$\delta\phi_{star}$. The dashed lines show the values $\langle\Delta\phi_{star}\rangle\pm\delta\phi_{star}$.}
\label{fig3}
\end{figure}
%
%Fig4
\begin{figure}
\centering
\includegraphics[width=9.5cm, keepaspectratio]{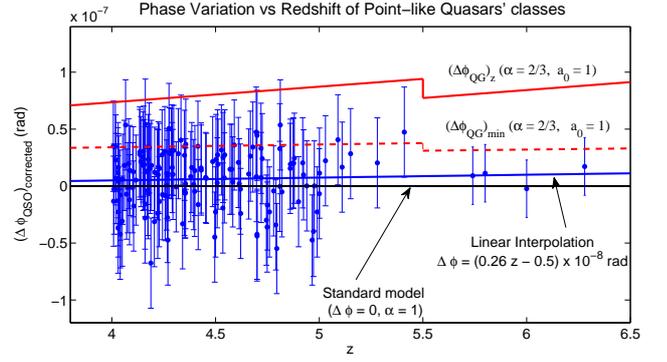}
\caption{Corrected phase variations of QSOs as a function of the redshift z.
No blurring effects due to QG are observed.
The linear interpolation of all the data sets show a positive trend, indicating the possible resolvability of the sources - blue solid line (color online).
The red (color online) continuous and dashed ``z''-shaped lines represent the expected growth of the phase variation, $\Delta\phi$, as a function of the redshift $z$. The break of the two curves indicating the limits of QG effects with (red solid line)  and without cosmological redshift corrections (red dashed line), are due to the change of filters, F775W and F850LP.}
\label{fig4}
\end{figure}
%

%% Fig1
In Fig.~\ref{fig3} we plot the phase variation of a sample of stars chosen with
the values of the maximum of the diffraction pattern in the same range as the QSOs. 
We indicate with $\langle\Delta\phi_{star}\rangle$ 
the averaged value of the phase variation of the sample at which we associate the error $\delta\phi_{star}$ defined as the
maximum difference between the phase variation values of the stars and $\langle\Delta\phi_{star}\rangle$. 

In Fig.~\ref{fig4} we plot the corrected phase variation of the QSOs, $(\Delta \phi(z)_{QSO})_\mathrm{Corrected}$,
at which we associate the error $\delta\phi_{star}$.
We found no blurring effects of QGs. The phase variations observed are much smaller than those expected from the space-time fluctuations described by the holographic principle model having $\alpha=2/3$ and $a_0=1$.
The phase variation $(\Delta \phi(z)_{QSO})_\mathrm{Corrected}$ and the 
redshift $z$, follow, at the first order, a linear relationship $(\Delta \phi(z)_{QSO})_\mathrm{Corrected}=az+b$, 
whose coefficients are $a=(0.26\pm 0.78)\times 10^{-8}$ and $b=(-0.5\pm 3.5)\times 10^{-8}$, 
indicating the possible resolvability of QSOs and/or their host galaxies.We make use of the relations
\begin{eqnarray}
 \sigma_a=\delta \phi_{star} \sqrt{\frac{\sum_i z^2_i}{\Delta}} \,\,\,\, \mathrm{and }\,\,\, \sigma_b=\delta \phi_{star} \sqrt{\frac{N}{\Delta}}
\end{eqnarray}
to evaluate the errors on the coefficients $a$ and $b$, where $N$ is the total number of QSOs and $\Delta=N\sum z_i^2-(\sum z_i)^2$.

\begin{figure}
\centering
\includegraphics[width=9.5cm, keepaspectratio]{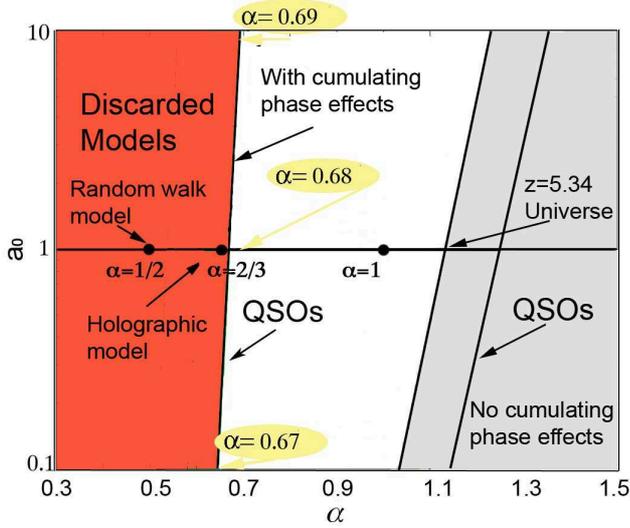}
\caption{Parameter space $a_0$ and $\alpha$ The red region (colour online) in the left is the portion of the parameter space that is discarded by our observations with the most conservative approach that takes in consideration the cumulating effects of space-time fluctuations (see text). The white region in the middle is the additional portion of the parameter space that would be discarded if no compensation due to the cumulating effect of space-time occurs.
The grey region in the right is the portion of the parameter space that cannot be excluded in any case.
The data obtained with our analysis from the farthest QSOs exclude the random walk and the holographic models of space-time foam in the range of $a_0$, here reported. These results, indicated by the label QSOs, go far beyond the previous estimates reported in the literature  (indicated by z=5.34 Universe), obtained without considering the cumulating effects of space-time.}
\label{fig5}
\end{figure}

To put our limits to the two parameters of the QG, $\alpha$ and $a_0$, we assume that
$(\Delta \phi(z)_{QSO})_\mathrm{Corrected}\geq \Delta \phi(z)_{QG}$. Our limits 
are conservative because in the quantity $(\Delta \phi(z)_{QSO})_\mathrm{Corrected}$ 
we cannot disentangle the QG effects from those from the resolvability of the QSOs. 
In Fig.~\ref{fig5}, we compare our results with the upper estimates found in the literature, which refer to the estimates obtained with sources located at distances up to $z=5.34$ (Ragazzoni et al., 2003).
 
\begin{figure}
\centering
\includegraphics[width=9.5cm, keepaspectratio]{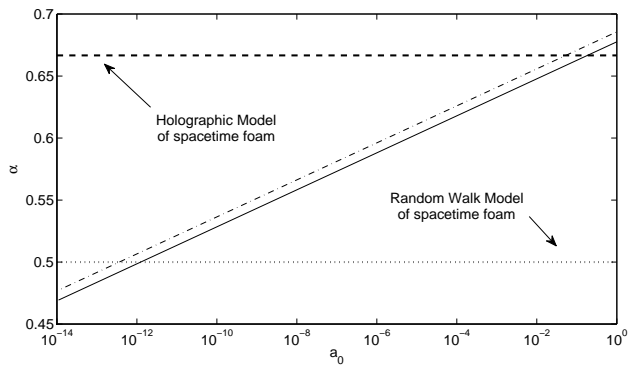}
\caption{Limitations to the parameters $a_0$ and $\alpha$ from QSOs. The excluded regions are below the dotted-dashed or the solid line depending on whether the photons redshifted to the observer wavelenghts, were taken into account.  The holographic model ($\alpha=2/3$) of space-time foam is preserved 
for $a_0 \lesssim 5~10^{-2}$ and $a_0 \lesssim 0.15$ (upper panel).
The random walk model ($\alpha=1/2$) is preserved  for $a_0 \lesssim 10^{-13}$ and  $a_0 \lesssim 10^{-12}$ (lower panel).}
\label{fig6}
\end{figure}

As reported in Fig.~\ref{fig6}, our results exclude the random walk model and most of the holographic scenario
of the space-time foam. For $a_0=1$, one obtains
$(\Delta \phi(z)_{QSO})_\mathrm{Corrected}=(0.26 \,z- 0.5)\times 10^{-8}$~rad.  
By using  the results previously discussed and the errors
on the estimate of the coefficients $a$ and $b$, one obtains the lower limit $\alpha_{min}=0.677$, with an $83.3 \%$ CL lower limit
without considering the expansion of the universe, and $\alpha_z=0.684$ with an $85.5 \%$ CL lower limit,
if we take the cosmological redshift of the photons emitted by QSOs to the observer's wavelength into account.
Both the results exclude the Holographic scenario, but for low values of the the lowest-order amplitude of the quantum fluctuations, $a_0 \lesssim 5\times10^{-2}$ and $a_0 \lesssim 0.15$. The expected value for the amplitude of the quantum fluctuation is $a_0\sim 1$ (Ng, 2003).

%% Fig5

%% Fig6

\section{Conclusions}

From the analysis of the diffraction images of the most distant QSOs, obtained with HST, we found 
no blurring effect that could be caused by the interaction of photons with quantum gravity fluctuations. 
Our results indicate that most of the models of space-time foam consistent with the holographic principle 
and, consequently, with black hole entropy, can be ruled out for most of the values of the fluctuation amplitude parameter $a_0$.

Surprisingly, we did not find any indirect trace of the magnification of QSO images expected from the geometry of the Universe adopted in our calculations that could generate a possible additional blurring in the diffraction images combined with that expected from QG effects. 
The linear regression found in our estimates indicates that an eventual combination of  QG effects and that possible of QSO magnification does not show any detectable contribution with the dataset available. If a significative trend were found, we should have estimated the QSO luminosities and calculated a QSO luminosity function (Richards et al., 2006)  to disentangle this contribution from the one of QG fluctuations.
One can suppose that either the Universe is flat in the interval $4<z<6$ or that QSOs may present some unknown evolutionary effects at large $z$.
Future observations with EELT, when equipped with adaptive optics, of QSOs at higher redshift also imaged with a filter at lower wavelengths should improve our results.

Finally, our results show the absence of any directional dependence on QG effects, since QSOs are located in very different regions of the sky at redshifts between $4<z<6.3$. This would also limit the
effects of intergalactic medium on the light across these cosmological distances in those observation bands. This confirms the cosmological principle on large scales. The isotropy on large scales of the Universe, verified by the distribution of superclusters of galaxies and QSOs on very large scales, is also verified through the independence of the behavior on smaller scales of the quantum fluctuations from any direction in space.
Viewed on a large scale, the properties of the Universe and the laws of physics are the same for all observers.

\subsection{The effect of QG fluctuations in quantum experiments}
We  can also infer from our results that quantum states will be preserved for very long distances because they are not affected by the blurring of space-time fluctuations. 
This is a very important result for quantum information and the foundations of quantum physics: a photon can travel undisturbed for billions of light years within the errors dictated by current HST technology. One can refer to \cite{kok03}, \cite{zeibook} and \cite{zeibook2} for deeper insight.
With our results we can set an upper limit to the phase factor imposed by the foamy structure of space-time to the photons that travel from the farthest QSOs caused by QG fluctuations, which is on the order of $\delta \phi \sim 2~10^{-8}$.

We consider an ideal quantum communication experiment, where Alice on the Earth wants to exchange single quanta of light with Bob located on the most distant QSO observed by HST, with wavelength  $\lambda=1 \, \mu$m and build a single--photon cryptographic quantum key.
The two parties will find that the probability of having one qubit mismatch caused by quantum gravity 
is almost negligible with respect to the errors usually present in those experiments, affecting at 
maximum only one qubit over a set of $\sim 5\times10^{7}$ qubits used in the building of the key. 
Earth--Moon and Earth--Mars experiments would cumulate a phase shift $\Delta \phi \sim 4.18\times10^{-14}$ 
and $\Delta \phi \sim 2.15\times10^{-13}$, respectively. 
The value $\alpha=1$ suggests a lower limit $\Delta \phi \sim 6.15\times10^{-17}$ when noncumulating 
QG fluctuations occur on the Planck scales. Similar argumentations are valid for entangled photons.

\section{Acknowledgments}
The authors would thank B. Bassett, Bo Thid\'e, M. Stiavelli and the anonymous referee for the useful comments and improvements to the work. One of us, FT, gratefully acknowledges the support of CARIPARO Foundation within the 2006 Program of Excellence.

\end{document}